\documentclass[a4paper,preprint, pre,showpacs,footinbib,superscriptaddress]{revtex4}
\usepackage{amsmath,amssymb}
\usepackage{makeidx}
\usepackage{amsfonts}
\usepackage[ansinew]{inputenc}
\usepackage[usenames,dvipsnames]{pstricks}
\usepackage{epsfig,graphicx}
\usepackage{auto-pst-pdf}

\usepackage{amsthm}
\newtheorem{theorem}{Theorem}
\newtheorem{lemma}{Lemma}

\newcommand{\ttheta}{{\theta_0}}
\newcommand{\htheta}{{\hat{\theta}}}

\newcommand{\rarrP}{{\;\xrightarrow{\mathrm{Prob.}}\;}}

\makeindex
\begin{document}

 \title{Consistent inference of a general model using the pseudo-likelihood method } 
\author{Alexander Mozeika}
\author{Onur Dikmen}
\affiliation{Department of Information and Computer Science, Aalto University, P.O. Box 15400, FI-00076,  Finland.}
\author{Joonas Piili}
\affiliation{Department of Biomedical Engineering and Computational Science, Aalto University, P.O. Box 12200, FI-00076, Finland}

\date{\today}

\begin{abstract}
Recently maximum pseudo-likelihood (MPL) inference method has been successfully applied to statistical physics models with intractable likelihoods. We use information theory to derive a relation between the pseudo-likelihood and likelihood functions. Furthermore, we show consistency of the pseudo-likelihood method for a general model.
\end{abstract}


\maketitle
 
%
As statistical physics (SP) models started to be widely used not only in its traditional domain but also to describe biological, financial, etc. phenomena, inferring  the interactions (model parameters) from the data has become an important research topic in the physics community~\cite{schneidman2006weak, Morcos2011direct, mastromatteo11criticality,  bialek12statistical}. This also strengthens a connection between SP and the areas of statistics and machine learning where model parameters are inferred from data. 
In statistics the maximum likelihood (ML) method is a standard approach due to its attractive statistical properties such as consistency, i.e. its ability to recover true parameters of a model, and asymptotic efficiency~\cite{ferguson1996course}. Unfortunately, a direct application of this method is usually infeasible  as it involves computing  the normalisation constant of a distribution (partition function in SP) which is nontrivial even for highly stylised  models of SP~\cite{Barahona1982complexity}.

For equilibrium Gibbs-Boltzmann distribution~\footnote{ The Gibbs-Boltzmann distribution is given by $P(s)=\mathrm{e}^{-\beta E(s)}/Z$, where $\beta$ is inverse temperature, $E(s)$ is an energy function and $Z=\sum_s\mathrm{e}^{-\beta E(s)}$ is a partition function. } the  ML method is usually implemented by so-called Boltzmann learning~\cite{ackley85learning} which uses samples from the distribution to approximate the gradients of the likelihood. However, it uses Markov chain Monte Carlo (MCMC) for sampling which can have very long equilibration times even for moderate system sizes. The sampling can be approximated for speed-up by stopping the MCMC early thus leading to contrastive divergence~\cite{carreira_contrastive} learning method. This method is very efficient but it is biased and not consistent~\cite{carreira_contrastive}. Other methods,  such as mean-field approximation~\cite{opper2001advanced}, allow to avoid the MCMC sampling but their statistical properties are generally not 
known.  

In the MPL method~\cite{besag75statistical} one avoids computation of a partition function by replacing the likelihood by a much simpler function of model parameters. Recently this method was successfully used for the protein contact prediction~\cite{ekeberg13improved} and it seems to outperform other methods for the benchmark  Ising-spin  models~\cite{aurell12inverse}. Furthermore, MPL method was shown to be consistent for the Ising model used in Boltzmann  learning~\cite{hyvarinen_consistency} and for the Gibbs-Boltzmann distributions over $\mathbb{Z}^d$~\cite{gidas1988consistency}. 

In this paper, we consider ML and MPL methods of inference. We show that both methods are equivalent to the problem of minimisation of a relative entropy between the distributions of model and  data. This has been  known for the ML method but for the MPL method this result is new. We use this framework to derive a relation between the likelihood and pseudo-likelihood functions. Furthermore, we prove consistency of MPL method for a general model.


%

Let us consider the following inference problem: we are given $L$ samples $\{s^\mu\}_{\mu=1}^L$ drawn independently from the  probability distribution $P_{\ttheta}(s)$, where $s = (s_1, s_2, ..., s_N)$, and we are required to estimate the \textit{true} parameters $\ttheta$ of this distribution. A classical approach to this problem is to maximise the log-likelihood~\footnote{This is equivalent to maximising the likelihood function $\prod_{\mu=1}^L P_\theta(s^\mu)$.} with respect to the parameters for given data
\begin{align}\label{equ:mle}
&\htheta_L = \arg\max_\theta \mathcal{L}_L(\theta)\\
&\mathcal{L}_L(\theta) = \frac1L \sum_{\mu=1}^L \log P_\theta(s^\mu).\nonumber
\end{align}
The  (ML) \textit{estimator} $\htheta$ obtained by the above procedure is \textit{weakly consistent} (resp. \textit{strongly consistent}):  in the large sample limit $L\rightarrow\infty$ we have that $\htheta\rightarrow\ttheta$ in probability (resp. almost surely) for all possible true values of $\ttheta$~\cite{ferguson1996course, Newey1994}.

With an infinite amount of data ($L=\infty$) the ML procedure (\ref{equ:mle}) allows us to find its true parameters $\ttheta$. To show this we will consider the difference
\begin{align}\label{equ:ml_cons}
&\frac1L  \sum_{\mu=1}^L \left( \log \hat{P}_L(s^\mu) - \log P_\theta(s^\mu)\right) \\
&= \sum_s \hat{P}_L(s) \log \frac{\hat{P}_L(s)}{P_\theta(s)}\nonumber\,=D(\hat P_L||P_\theta),
\end{align}
where $\hat{P}_L(s)=\frac1L \sum_{\mu=1}^L \delta_{s,s^\mu}$, with $\delta_{s,s^\mu}$ denoting Kronecker delta function, is an \textit{empirical} distribution of data. Thus the maximisation of log-likelihood in \eqref{equ:mle} is equivalent to the minimisation of the  function  $D(\hat P_L||P_\theta)$  which is a \textit{relative entropy} (or Kullback-Leibler divergence) of Information theory~\cite{cover2012elements}. By the  Strong Law of Large Numbers we have that $\lim_{L\rightarrow\infty}\sum_s \hat{P}_L(s) \log \frac{\hat{P}_L(s)}{P_\theta(s)}=D(P_\ttheta||P_\theta)$, where $D(P_\ttheta||P_\theta) = \sum_s P_\ttheta(s) \log \frac{   P_\ttheta(s)}{P_\theta(s)}$.  Note that $D(P_\ttheta||P_\theta) \geq 0$ with equality if and only if $P_\ttheta(s) = P_\theta(s)$ holds for all $s$~\cite{cover2012elements}. Furthermore, assuming that the equality of distributions $P_\ttheta(s) = P_\theta(s)$ implies the  equality  of its parameters $\ttheta=\theta$ (this is so-called \textit{identifiability} condition) 
completes the  proof. We note that if the limit and maximisation operators in $\lim_{L\rightarrow\infty}\htheta_L= \lim_{L\rightarrow\infty}\arg\max_\theta \mathcal{L}_L(\theta)$ \textit{commute} then the above argument also shows (strong) consistency of the ML estimator $\htheta_L$. This requirement imposes furter conditions on the estimator function $\mathcal{L}_L(\theta)$~\cite{ferguson1996course, Newey1994}.  

%
Although the ML estimator is consistent, very often the method of inference itself  is not practical as it requires the computation of the partition function~\footnote{For e.g. the partition function $Z=\sum_s \mathrm{e}^{-\beta E(s)}$, where  $s\in\{-1,1\}^N$,  of Ising model is a sum over $2^N$ states.}. One of the ways to circumvent this problem is instead of a log-likelihood to  maximise a much simpler pseudo-log-likelihood 
\begin{align}\label{equ:logpseudo}
&\htheta_L = \arg\max_\theta \mathcal{PL}_L(\theta)\\
 &\mathcal{PL}_L(\theta) = \frac1L \sum_{\mu=1}^L \sum_{i=1}^N \log P_\theta(s_i^\mu|s_{-i}^\mu)\nonumber,
\end{align}
where $P_\theta(s_i|s_{-i})=P_\theta(s)/\sum_{s_i} P_\theta(s)$, with $s_{-i}=(s_1,\ldots,s_{i-1},s_{i+1},\ldots,s_N)$, is a conditional distribution. The conditional distribution is, by definition, independent of the partition function.  As in the case of log-likelihood the pseudo-log-likelihood method \eqref{equ:logpseudo} is also equivalent  to the minimisation of a relative entropy.  This can be shown as follows.

Firstly, using the relative entropy  \eqref{equ:ml_cons} and  equality  $P_\theta(s)=P_\theta(s_i|s_{-i})P_\theta(s_{-i})$,  which is true for each $i$,  we obtain 
\begin{align}\label{equ:ml_cons2}
N D(\hat{P}_L||P_\theta) &= N \sum_s \hat{P}_L(s) \log \frac{\hat{P}_L(s)}{P_\theta(s)}\\
&= \sum_{i=1}^N \sum_s \hat{P}_L(s) \log \frac{\hat{P}_L(s)}{P_\theta(s_i|s_{-i})P_\theta(s_{-i})}.\nonumber
\end{align}
Let us  now  in the above replace the distribution of model $P_\theta(s_{-i})$ by  the empirical distribution of data $\hat{P}_L(s_{-i})$. This gives rise to the new probability distribution $\hat{P}_{\theta\vert\theta_0}^i(s)=P_\theta(s_i|s_{-i})\hat{P}_L(s_{-i})$ and immediately leads us to the inequality 
\begin{align}\label{equ:kl_pseudo}
\sum_{i=1}^N \sum_s \hat{P}_L(s) &\log \frac{\hat{P}_L(s)}{P_\theta(s_i|s_{-i})\hat{P}_L(s_{-i})} \nonumber\\
& = \sum_{i=1}^N D(\hat{P}_L||\hat{P}_{\theta\vert\theta_0}^i) \geq 0\,.
\end{align}
Clearly the minimum of the above sum of relative entropies, with respect to the model parameters $\theta$, corresponds to the maximum of the  pseudo-log-likelihood function used in \eqref{equ:logpseudo}. This sum is also a lower bound for the (rescaled by $N$) relative entropy \eqref{equ:ml_cons2}.  In order to show this we consider the difference 
\begin{align}\label{equ:bound}
N D&(\hat{P}_L||P_\theta) - \sum_{i=1}^N D(\hat{P}_L||\hat{P}_{\theta\vert\theta_0}^i)\\
= &  \sum_{i=1}^N \sum_s \hat{P}_L(s) \ln{\frac{\hat{P}_L(s_{-i})}{P_\theta(s_{-i})}} \nonumber\\
= & \sum_{i=1}^N \sum_{s_{-i}} \hat{P}_L(s_{-i}) \ln{\frac{\hat{P}_L(s_{-i})}{P_\theta(s_{-i})}}\geq 0.\nonumber
\end{align}
The inequality in the above is due to the last line being a sum of relative entropies. 

A consequence of the inequality \eqref{equ:bound} is the relation 
\begin{align}\label{equ:ineq}
\mathcal{PL}_L(\theta) -\sum_{i=1}^N H_i(\hat{P}_L) \geq   N\mathcal{L}_L(\theta), 
\end{align}
where $H_i(\hat{P}_L)=- \sum_{s_{-i}}\hat{P}_L(s_{-i})\ln \hat{P}_L(s_{-i})$ is a Shannon entropy of the empirical distribution $\hat{P}_L(s_{-i})= \sum_{s_i}\hat{P}_L(s)$, between the objective functions of ML \eqref{equ:mle} and MPL  \eqref{equ:logpseudo} methods. 
%
%
 Furthermore, using the inequality \eqref{equ:kl_pseudo}  we can show that the MPL procedure (\ref{equ:logpseudo}) recovers the true parameters   $\theta_0 $ with an infinite amount of data. To show this we consider the sum of relative entropies
\begin{align}
\sum_{i=1}^N D(P_\ttheta||P_{\theta\vert\theta_0}^i) &= \sum_{i=1}^N \sum_s P_\ttheta(s) \log \frac{P_\ttheta(s)}{P_\theta(s_i|s_{-i}) P_\ttheta(s_{-i})}\nonumber\\
&= \sum_{i=1}^N \sum_s P_\ttheta(s) \log \frac{P_\ttheta(s_i|s_{-i}) P_\ttheta(s_{-i})}{P_\theta(s_i|s_{-i}) P_\ttheta(s_{-i})}\nonumber\\
&= \sum_{i=1}^N \sum_s P_\ttheta(s) \log P_\ttheta(s_i|s_{-i}) - \sum_{i=1}^N \sum_s P_\ttheta(s) \log P_\theta(s_i|s_{-i})\nonumber\\
&= Q_0(\ttheta) - Q_0(\theta) \geq 0\,,
\end{align}
where $Q_0(\theta)= \lim_{L\rightarrow\infty}\mathcal{PL}_L(\theta)$. Thus, $Q_0(\theta) \leq Q_0(\ttheta)$ and if $P_\ttheta(s_i|s_{-i}) \neq P_\theta(s_i|s_{-i})$ implies that $\ttheta\neq\theta$ then $\theta=\ttheta$ is the unique maximum of $Q_0(\theta)$. We note that this proves the condition i) of the Theorem \ref{theorem:1} in the Appendix \ref{section:appendix}. We will use this theorem to show (weak) consistency of the MPL estimator (\ref{equ:logpseudo}).

Let us assume that $\theta\in \Theta$, where $\Theta $ is a compact set (this is the condition ii) of the Theorem \ref{theorem:1}) and define $\hat{Q}_L(\theta) =\mathcal{PL}_L(\theta)$. If $Q_0(\theta)$ is a continuous function of $\theta$ and $\hat{Q}_L(\theta)$ converges uniformly in probability to $Q_0(\theta)$, i.e. $\sup_{\theta \in \Theta}\left\vert \hat{Q}_L(\theta)- Q_0(\theta)\right\vert \rarrP 0 $ as $L\rightarrow\infty$,  then  the conditions iii) and iv) of the Theorem \ref{theorem:1} are satisfied.  In order to prove these conditions we will use the Lemma \ref{lemma:1} in the Appendix \ref{section:appendix}. To this end we define $q(s,\theta) = \log \prod_{i=1}^N P_\theta(s_i|s_{-i})$ and hence $\hat{Q}_L(\theta)=(1/L)\sum_{\mu=1}^L q(s^\mu,\theta)$. Now let us assume that the function $q(s,\theta)$ is continuous at each $\theta\in\Theta$ and consider  
\begin{align}\label{equ:ub}
\vert q(s,\theta)\vert &= \left\vert\log \prod_{i=1}^N P_\theta(s_i|s_{-i})\right\vert
\leq \sup_{\theta\in\Theta} \left\vert \log \prod_{i=1}^N P_\theta(s_i|s_{-i})\right\vert = d(s)\,,
\end{align}
then if $\sum_s P_\ttheta(s) d(s) < \infty$,  we have that $Q_0(\theta)$ is continuous and  $\hat{Q}_L(\theta)$ converges uniformly in probability to $Q_0(\theta)$ by Lemma \ref{lemma:1}. Thus if (a) $P_\ttheta(s_i|s_{-i}) \neq P_\theta(s_i|s_{-i})$ implies that $\ttheta\neq\theta$;  (b) $\theta\in \Theta$, where $\Theta $ is a compact set; (c) $\log \prod_{i=1}^N P_\theta(s_i|s_{-i})$ is continuous; (d) $\sum_s P_\ttheta(s)\sup_{\theta\in\Theta} \left\vert \log \prod_{i=1}^N P_\theta(s_i|s_{-i})\right\vert  < \infty$, then the MPL estimator (\ref{equ:logpseudo}) is weakly consistent, i.e. $\hat{\theta}\rarrP \ttheta$ as  $L\rightarrow\infty$.


To summarise,  we mapped the maximum likelihood (ML) and maximum pseudo-likelihood (MPL) methods of inference onto information theory framework which allows us to investigate the relation between these two methods. In this framework for both methods the relative entropy is an objective function,  minimisation of which is equivalent to ML and MPL. Furthermore, we derive an inequality which establishes a relation between the likelihood and pseudo-likelihood functions. Finally, we prove (weak) consistency of pseudo-likelihood method for a general probability distribution. We envisage that the strong consistency of MPL can also be proven by, for example, adopting the consistency proof of ML  in~\cite{ferguson1996course}. Also,  all derivations in this paper are for the distributions of discrete variables but we expect that extending these results to the case of continuous variables is a straightforward matter. 

\begin{acknowledgments} 
The work of Alexander Mozeika (AM) and Onur Dikmen (OD) is supported by the Academy of Finland (Finnish Centre of Excellence in Computational Inference Research COIN, 251170). AM was also supported in part by the Academy of Finland, grant 256287. The work of Joonas Piili is supported by The Emil Aaltonen Foundation. AM is thankful to  Manfred Opper for interesting and helpful discussions. AM and OD would like to thank R\'{e}mi Lemoy for discussions. 
\end{acknowledgments}


\appendix
\section{Theorem and Lemma\label{section:appendix}}
Here we state the Theorem  and Lemma (on the page 2121 and 2129 of~\cite{Newey1994} respectively)  which are used in the main text.

\begin{theorem}
\label{theorem:1}
If there is a function $Q_0(\theta)$ such that\\
i) $Q_0(\theta)$ is uniquely maximized at $\ttheta$;\\
ii) $\theta\in \Theta$, where $\Theta $ is a compact set;\\
iii) $Q_0(\theta)$ is continuous;\\
iv) $\hat{Q}_L(\theta)$ converges uniformly in probability to $Q_0(\theta)$,\\
then $\hat{\theta}\rarrP \ttheta$ as  $L\rightarrow\infty$.
\end{theorem}
\begin{lemma}
\label{lemma:1}
If $s^\mu$, where  $\mu=1,\ldots, L$,  are drawn independently from the probability distribution $P(s)$; $\Theta$ is a compact set; $q(s^\mu,\theta)$ is continuous at each $\theta\in\Theta$ with probability one; there is $d(s)$ with $\vert q(s,\theta)\vert \leq d(s)$ for all $\theta\in\Theta$ and $\sum_s P(s) d(s)<\infty$, then $\sum_s P(s) q(s,\theta)$ is continuous and
\begin{align*}
\sup_{\theta\in\Theta}\left\vert \frac{1}{L}\sum_{\mu=1}^L q(s^\mu,\theta) - \sum_s P(s) q(s,\theta)\right\vert \rarrP 0\,
\end{align*}
as  $L\rightarrow\infty$.
\end{lemma}
%



\end{document}